\documentclass[%
 reprint,
superscriptaddress,
 amsmath,amssymb,
 aps,
]{revtex4-2}

\usepackage{float}
\usepackage{siunitx}
\usepackage{xcolor}
\usepackage{graphicx}
\usepackage{dcolumn}
\usepackage{bm}
\usepackage{hyperref}

\begin{document}

\preprint{APS/123-QED}
\title{Spin polarization of Quantum Hall states for filling factors $1\le\nu\le2$\\measured with microcavity polaritons}
\author{Odysseas Williams}
\affiliation{Laboratory for Solid State Physics, ETH Zürich, 8093 Zürich, Switzerland}

\author{Stefan Faelt}
\affiliation{Laboratory for Solid State Physics, ETH Zürich, 8093 Zürich, Switzerland}
\affiliation{Institute of Quantum Electronics, ETH Zürich, 8093 Zürich, Switzerland}

\author{Filip Krizek}
\affiliation{Laboratory for Solid State Physics, ETH Zürich, 8093 Zürich, Switzerland}
\affiliation{IBM Research Europe-Zurich, 8803 Rüschlikon, Switzerland}

\author{Werner Wegscheider}
\affiliation{Laboratory for Solid State Physics, ETH Zürich, 8093 Zürich, Switzerland}

\date{\today}
\begin{abstract}
Spin polarization measurements were performed in three 2D Electron Gases in GaAs with densities $n_e =$ 9.1, 7.2 and 6.5 $\times 10^{10}$ cm$^{-2}$, in the quantum Hall regime. Full spin polarization at $\nu=1$ surrounded by rapid depolarization due to Skyrmion formation was observed in all devices, consistent with past measurements. Depolarization of the $\nu=4/3$, 8/5 states and repolarization of the $\nu=5/3$ state was also measured, in remarkable agreement with a non-interacting, disorder-free Composite Fermion model. Optical power and temperature dependent measurements of the $\nu=1$ state suggest a regime of non-linear optics. 
\end{abstract}
\maketitle


\section{Introduction :\\
Cavity polaritons and the Quantum Hall effect}

When placing a 2-Dimensional Electron Gas (2DEG) with electron density $n_e$ in a perpendicular magnetic field $B$, the kinetic energy term of the electron states will morph into a set of energetically degenerate Landau Levels (LL), which can be semiclassically interpreted as all the electrons performing cyclotron orbits. The density of states for each degenerate LL per spin state is given by $n_{LL} = eB/h$ with $e$ the electron charge and $h$ Planck's constant. At strong enough magnetic fields, the condition $n_e = \nu\:n_{LL}$ is met, with $\nu$ an integer. 

At low enough temperatures and disorder, this means the 2DEG electrons are exactly split among the first $\nu$ LLs, giving rise to the now famous Integer Quantum Hall (IQH) effect \cite{Klitzing1958}. This discovery was soon followed by the addition of fractional values of $\nu$ giving rise to the similar Fractional Quantum Hall (FQH) effect \cite{Stormer1999}. While its existence has been theorised for some specific fractional values, first by Laughlin's trial wavefunction \cite{Laughlin1999}, and nowadays more widely used Composite Fermion (CF) model \cite{Jain1989}, its full origin, in particular for fractions in the $2<\nu<4$ range, remains an active topic of research.

While bulk electrical transport measurements of quantum Hall states remain the primary way of detecting them, localized measurements of the 2DEG spin polarization shed some light on fundamental aspects of these states, such as 2DEG self-interactions. Spin polarization of the $\nu=1$ state for instance, measured with optical absorption, showed compatibility with the formation of Skyrmions as excitations of the system \cite{Aifer1996,Manfra1996,Plochoka2009}. 

Other techniques involving Nuclear Magnetic Resonance (NMR) or spin-resolved pumped tunneling were used to measure spin polarization of FQH states \cite{Yoo2020,Smet2001,Tiemann2012}. While these techniques are less perturbative than optical absorption measurements, they still induce perturbations by forcing current through the 2DEG. Furthermore, they are performed and averaged over large sections of the 2DEG, meaning the fragile FQH states are more strongly affected by spatial disorder.

In this paper, we build upon previous spin polarization measurements of quantum hall states, realized with polariton spectroscopy \cite{Lupatini2020,Rav2018}. Cavity polaritons are light-matter hybrid excitations formed when trapping photons in a mirror cavity coupled to an excitable medium. They have been studied in a variety of materials \cite{Weisbuch1992,byrnes2014}. Coupling an optical cavity to a GaAs Quantum Well (QW) doped with a 2DEG has garnered some interest for allowing to engineer photon-photon interactions through their excitation counterpart interactions \cite{Masharin2023,Julku2021,Knueppel2019}. Moreover, these polaritons can be used as non-perturbative local probes of 2DEGs \cite{Gabbay2007,Smolka2014,Franchini2021}.

In the context of spin polarization measurements, cavity-polaritons have only been used to characterize the $\nu=1$ state. With improved devices and measurement procedures, we show a remarkable agreement of our measurements with a disorder-free, non-interacting CF model around filling factor $\nu=3/2$. These results demonstrate the accuracy and the limited perturbation induced by this technique, and pave the way for further investigation of the 5/2 or 7/2 states. 

A time-resolved 4-wave mixing experiment previously reported an enhanced nonlinear optical response of similar devices when placed into the $\nu=2/5$, 2/3 and 1 states \cite{Knueppel2019}. In this report, a large plateau of vanished coupling in the $\nu=1$ state is interpreted as a manifestation of electron incompressibility and nonlinear optics. This result, not reported before, suggests this new generation of devices could exhibit even stronger nonlinearities, bringing them closer to the alluring polariton blockade regime \cite{Delteil2019}.

\section{Experimental considerations}\label{sec:experiment}

\subsection{Device characteristics}

The heterostructures used in this work were grown using Molecular Beam Epitaxy (MBE) and consist of the 2DEG at the center of a planar microcavity. On each side of the cavity are Distributed Bragg Reflector (DBR) mirrors, formed by periodically stacking $\lambda/4$ width AlAs and Al$_{0.1}$Ga$_{0.9}$As layers, separated by a $k\lambda/2$ width layer, with $k$ an integer and $\lambda$ the resonant wavelength. The 23 nm wide GaAs QW is thereby placed at the highest amplitude anti-node of the electric field to maximize 2DEG excitations coupling with cavity photons. The modulation doping is provided from both sides by Si delta-doping at nodes of the electric field, $3\lambda/4$ away from the center of the QW at the cavity center. The Si doping is placed in 9 nm wide GaAs QWs (called doping QWs from here on) and positioned at cavity nodes with the purpose of minimizing changes in the doping efficiency by the light from the optical probe. 

They were used as grown, with no extra fabrication step and no tunable 2DEG density via the use of gate to preserve high 2DEG mobility. The precise growth parameters were determined with transfer matrix calculations to optimize the cavity $Q$-value and transmittance. Table \ref{tab:samples} summarizes the different characteristics of our devices. 

\begin{table}[h]
\caption{\label{tab:samples}Devices used in this experiment}
\begin{tabular}{|c||c|c|c|c|}
\hline
 Sample & \begin{tabular}{@{}c@{}} DBR periods \\ (top/bottom) \end{tabular} & Cavity & \begin{tabular}{@{}c@{}} $n_e$ \\ ($10^{10}$cm$^{-2}$) \end{tabular}& \begin{tabular}{@{}c@{}} $dn_e/d\log_{10}(W)$\\($10^{10}$cm$^{-2}$)\end{tabular}\\ 
 \hline
 \hline
 A & 16/19.5 & $2\lambda$ & $9.1$ & 0\\
 \hline
 B & 17/20.5 & $1\lambda$ & $7.2$ & $-0.2$\\
 \hline
 C & 20.5/24& $\lambda/2$ & $6.5$ & $-0.2$\\
 \hline
\end{tabular}
\end{table}

The main difference in design in between the three wafers is their central cavity, where wafers A and B have respectively a $2\lambda$ and $1\lambda$ length Al$_{0.1}$Ga$_{0.9}$As cavity, whereas wafer C has a $\lambda/2$ length AlAs cavity (albeit mixed with some GaAs and AlGaAs layers).

Since the narrow doping QWs also get populated with electrons (about 10 times more than the central 2DEG), the sheet resistance could not be accurately determined as the contact resistance to each layer for any contact was not identical. The fraction of the current that is going through the 2DEG versus the doping QWs could not be determined and with that, only a lower bound for the mobility could be established. The $n_e$ could however be deduced from their quantum Hall optical signatures (see Section \ref{subsec:LL pol sig}).

For potential nonlinear optics experiments, their light sensitivity, i.e. the variation of their density $n_e$ with incident power density $W$, was also an important characteristic. Only sample A showed complete light insensitivity (see Section \ref{subsec:power_temp_dep}).

\begin{figure}[h]  
    \includegraphics[width=3.4in]{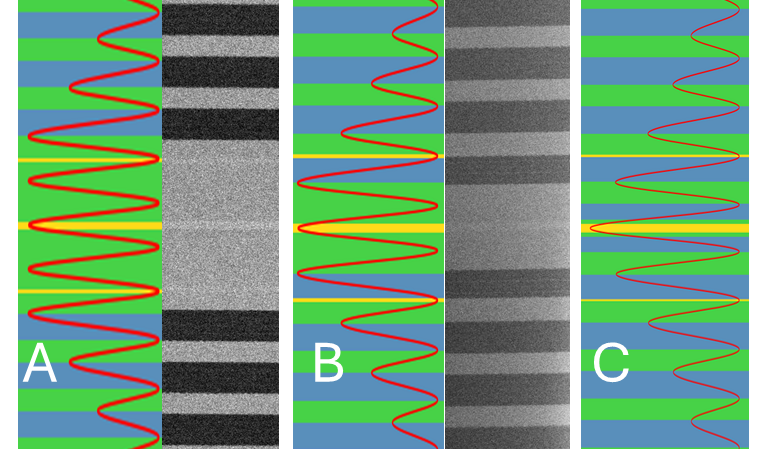}
    \caption{Simulations of middle parts of samples A, B and C (colored figures from left to right), corresponding respectively to a 2$\lambda$, 1$\lambda$ and $\lambda/2$ cavity. The blue layers are AlAs, the green layers are Al$_{0.1}$Ga$_{0.9}$As, and the yellow layers are the GaAs QWs. The intensity of light with wavelength $\lambda=815$ nm transmitted through the cavity was calculated with a transfer matrix method and plotted in red. The black and white images show Transmission Electron Microscope (TEM) images of samples A and B}\label{fig:TEM}  
\end{figure}

\subsection{Tuning the cavity wavelength}
\label{sec:cavity_tuning}

The epitaxially grown layers are thinner at the outer edges of the wafer. This is used to tune the cavity resonance by lateral position, where the cavity is red-shifted from the 2DEG resonance towards the center of the wafer and blue-shifted towards the edges.

A $5\times5$ mm piece was cleaved from each wafer and cooled down in a $^3$He cryostat to temperature $T=0.25$ K. The cryostat is fitted with a fiber-coupled scanning confocal microscope, through which a 800-850 nm spectral width LED beam was focused on a $\sim 10$ $\mu$m radius spot, which could be moved across the sample with piezoelectric nanopositioners. The reflected light was analyzed in a spectrometer and passed on to an algorithm to extract the reflection spectrum $R$ of the device.

The algorithm works by averaging over the top 10\% of recorded intensities for each wavelength bin when assembling all the spectra. This maximal intensity is mapped to a reflectance $R=1$, as we assume the high $Q$-value cavity to be very reflective away from the cavity mode.

\begin{figure}[h]  
    \includegraphics[width=3.4in]{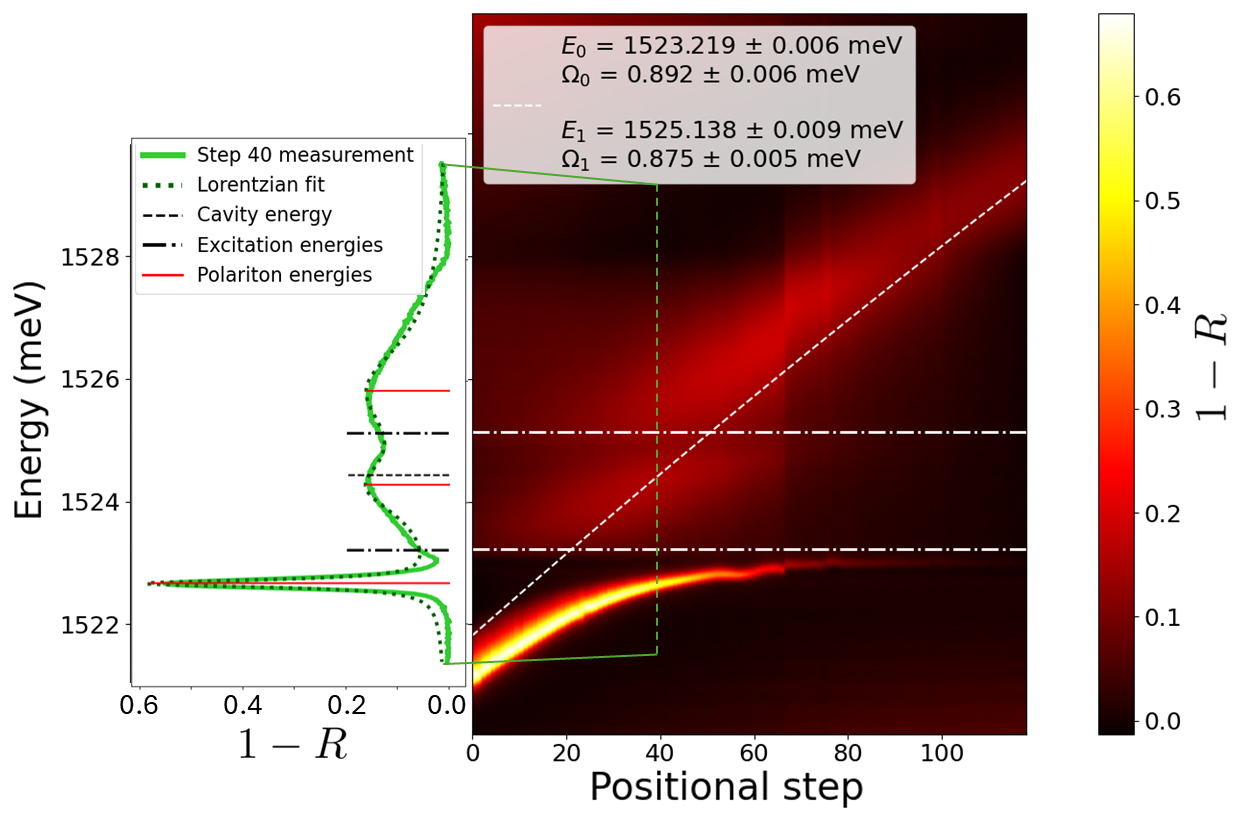}
    \caption{Right: Differential reflection spectra at varying positions on sample A. The diagonal white dashed line is the bare cavity energy, the horizontal dash-dotted lines are excitation energies of the central QW. Left: Differential reflection spectrum at positional step 40. The green dotted line are lorentzian functions fitted to extract the peak transmission energies. The black dashed and dash-dotted lines correspond to the white lines of the right side figure. The red lines are the polariton energies obtained by solving the coupled oscillator hamiltonian (Eq. \ref{eq:coupled_oscillator_ham}).}\label{fig:position_scan}  
\end{figure}

Figure \ref{fig:position_scan} shows differential reflection spectra $1-R$ of sample A at regularly spaced $x$ positions, as well as a single spectrum when the cavity energy is in between the two lowest excitations of the GaAs QW. 

The peak transmission energies $\mathcal{P}_0(x)$, $\mathcal{P}_1(x)$ and $\mathcal{P}_2(x)$, corresponding to polariton modes, were extracted by fitting Lorentzian distributions to the spectra. The QW excitation energies, $E_0$ and $E_1$, and their coupling energy to cavity photons (also known as Rabi coupling), $\Omega_0$ and $\Omega_1$, could then be calculated by fitting a coupled oscillator Hamiltonian,
\begin{equation}
\label{eq:coupled_oscillator_ham}
\hat{H}=
\begin{bmatrix}
E_0 & 0 & \Omega_0 \\
0 & E_1 & \Omega_1 \\
\Omega_0 & \Omega_1 & E_C(x)
\end{bmatrix},
\end{equation}
such that its eigenvalues correspond to the polariton modes \cite{Hopfield1958,Deng2010}. Here, $E_C(x)$ labels the bare cavity energy, and was approximated as a second order polynomial to match the sample thickness profile. Although we do not have direct readings of the spot positions, these become irrelevant if we assume our $E_C(x)$ fit to be accurate. 

As can be seen Figure \ref{fig:position_scan}, we observe multiple transmission energies when the cavity energy approaches resonance with the QW excitations, but only a single mode is observed away from it, corresponding roughly to the bare cavity transmission. The excitation energies $E_0$ and $E_1$ correspond respectively to the promotion of an electron from the Heavy Hole and Light Hole valence bands to the conduction band \cite{Lupatini2020}. Note that the lowest polariton branch is very bright and narrow compared to the higher order branches, which is due to scattering from the highest order branches to the lowest, as well as increased momentum dispersion of the higher order branches.

\subsection{Landau levels polariton signature}
\label{subsec:LL pol sig}
When a magnetic field $B$ is applied to a sample, the excitation energies shown in Figure \ref{fig:position_scan} split into multiple levels corresponding to the different LLs. At low fields, these levels are indistinguishable due to spectral overlap, as their cyclotron energies are smaller than their linewidths. The cyclotron energy for each LL is given by:
\begin{equation}
E_L = \left( L + \frac{1}{2} \right) \frac{\hbar e B}{m^{*}}
\end{equation}
where $\hbar$ is the reduced Planck constant, $m^{*}$ the electron effective mass and $L$ refers to the LL index (starting at 0). 

At low B fields, the lowest LLs are fully occupied due to their low degeneracy compared to the electron density, rendering them optically inactive. As the $B$ field strength increases, it separates the LLs and transitions across even filling factors $\nu$ are reached. When this occurs, a new LL with index $L = \nu/2 - 1$ becomes optically accessible. The factor 1/2 accounts for the two possible spin states in each LL. 

\begin{figure}[h]  
    \includegraphics[width=3.4in]{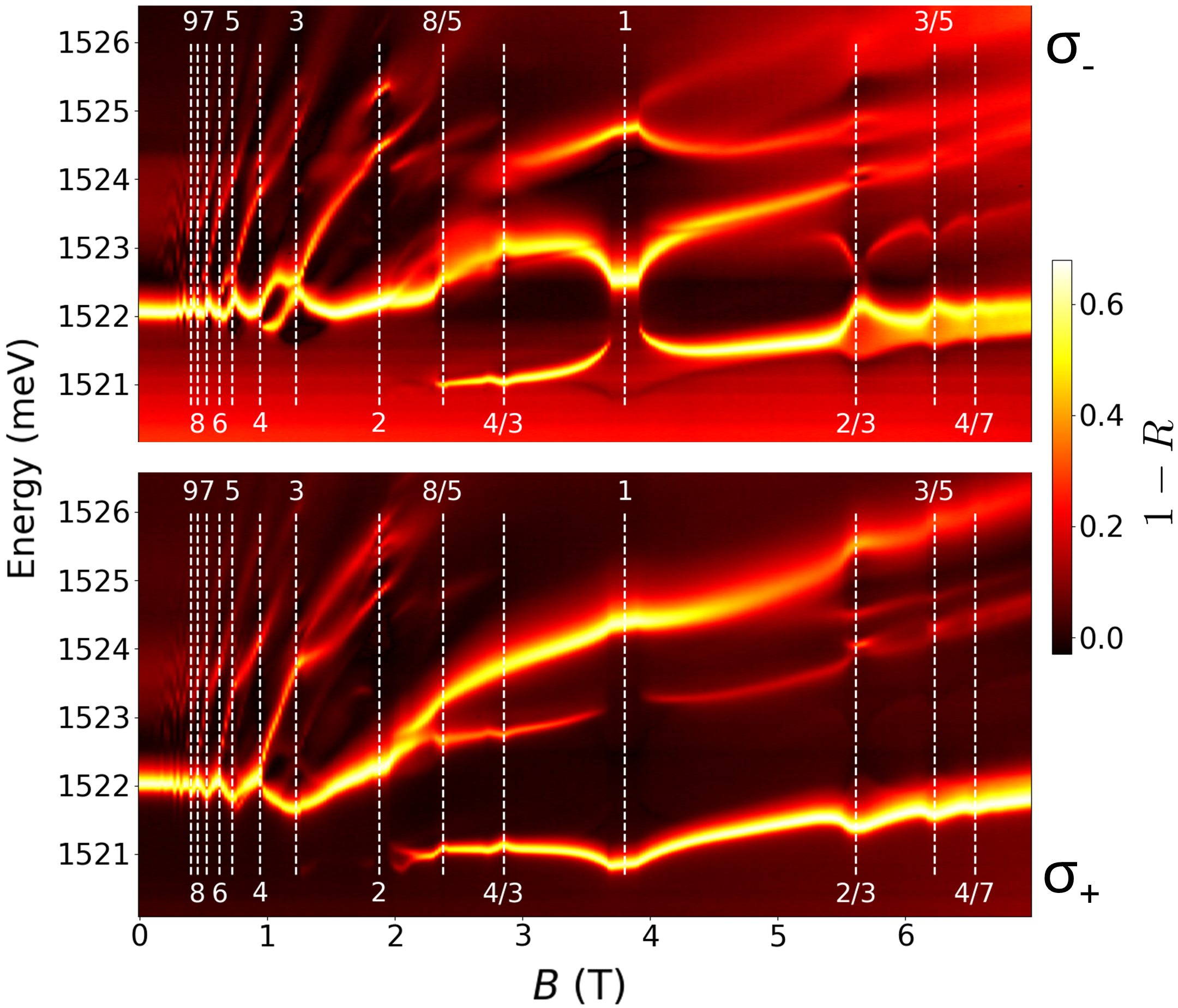}
    \caption{Differential reflection spectra of sample A as a function of applied $B$ field, for opposite optical circular polarization (top: $\sigma_-$ polarization, bottom: $\sigma_+$ polarization). The cavity energy is kept constant close to 1522 meV. The white dashed lines indicate integer and fractional filling factors $\nu$.}\label{fig:large_B_sweep}  
\end{figure}

When the cavity energy is tuned close to a newly available LL, we observe the single cavity mode splitting in two as $\nu$ is crossed. These two modes continue to separate as $B$ increases  (and $\nu$ decreases) due to the growing density of available states and, consequently, increased optical coupling strength.

Figure \ref{fig:large_B_sweep} illustrates this phenomenon. The cavity energy was tuned to 1522 meV, and measurements were taken using opposite circular polarizations ($\sigma_-$ and $\sigma_+$). At low $B$ fields, the transmission energy remains constant. As B increases, the cavity begins to couple with progressively lower LLs, causing the transmission energy to oscillate. At $B = 0.95$ T, corresponding to filling factor $\nu = 4$ (indicated by a white dashed line), the $L = 1$ level starts to empty. Since the cavity is tuned close to this transition, we observe the single transmission mode splitting into two distinct polariton branches as $L = 1$ empties.

Optical selection rules associate a specific photon circular polarization to excitations of different electron spin states in the QW \cite{Rav2018}. Starting from depolarized light producing multiple transmission peaks, we could select the circular polarization by adjusting our polarizer until some of the peaks were suppressed. When $\nu = 3$ is reached at $B = 1.27$ T, we observe different behaviors for $\sigma_-$ and $\sigma_+$ polarizations:
\begin{enumerate}
    \item For $\sigma_-$ polarization, the coupling initially increases from $\nu = 4$ but then decreases as $\nu = 3$ is approached, resulting in a very small splitting.
    \item For $\sigma_+$ polarization, the polariton coupling to $L = 1$ gradually increases from $\nu = 4$ to a maximum at $\nu = 3$.
\end{enumerate}
This difference arises because the circular polarizations couple to different spin states of the same LL transition. At $\nu = 3$, the $L = 1$ level is strongly spin-polarized, with one spin state nearly fully occupied and the other empty. We will explore this phenomenon in more detail in Section \ref{sec:results} for the range $2>\nu>1$.

As the $B$ field is further increased, an analogous behaviour is observed as $\nu=2$ and $\nu=1$ are reached (at 1.9 and 3.8 T respectively). The lowest polariton branch appears somewhat later than the $\nu=2$ state because the cavity is tuned higher than the $L=0$ transition. The lowest branch thus only appears once the optical coupling is very strong. The $\nu=1$ state again shows opposite behaviour of the lowest polariton for the two circular polarizations. Importantly, a similar effect is observed for other $B$ values corresponding to FQH states, namely $\nu=$ 4/7, 3/5, 2/3, 4/3, 8/5. By identifying all these filling factors, one can measure the 2DEG density of our devices. Samples B and C (not shown here) demonstrated the same behaviour.

\section{In depth measurements of IQH and FQH states}
\label{sec:results}

\subsection{Optical transitions to the $L=0$ level for opposite circular polarizations}
\label{subsec:optical transitions}

\begin{figure}[htb]  
    \includegraphics[width=3.4in]{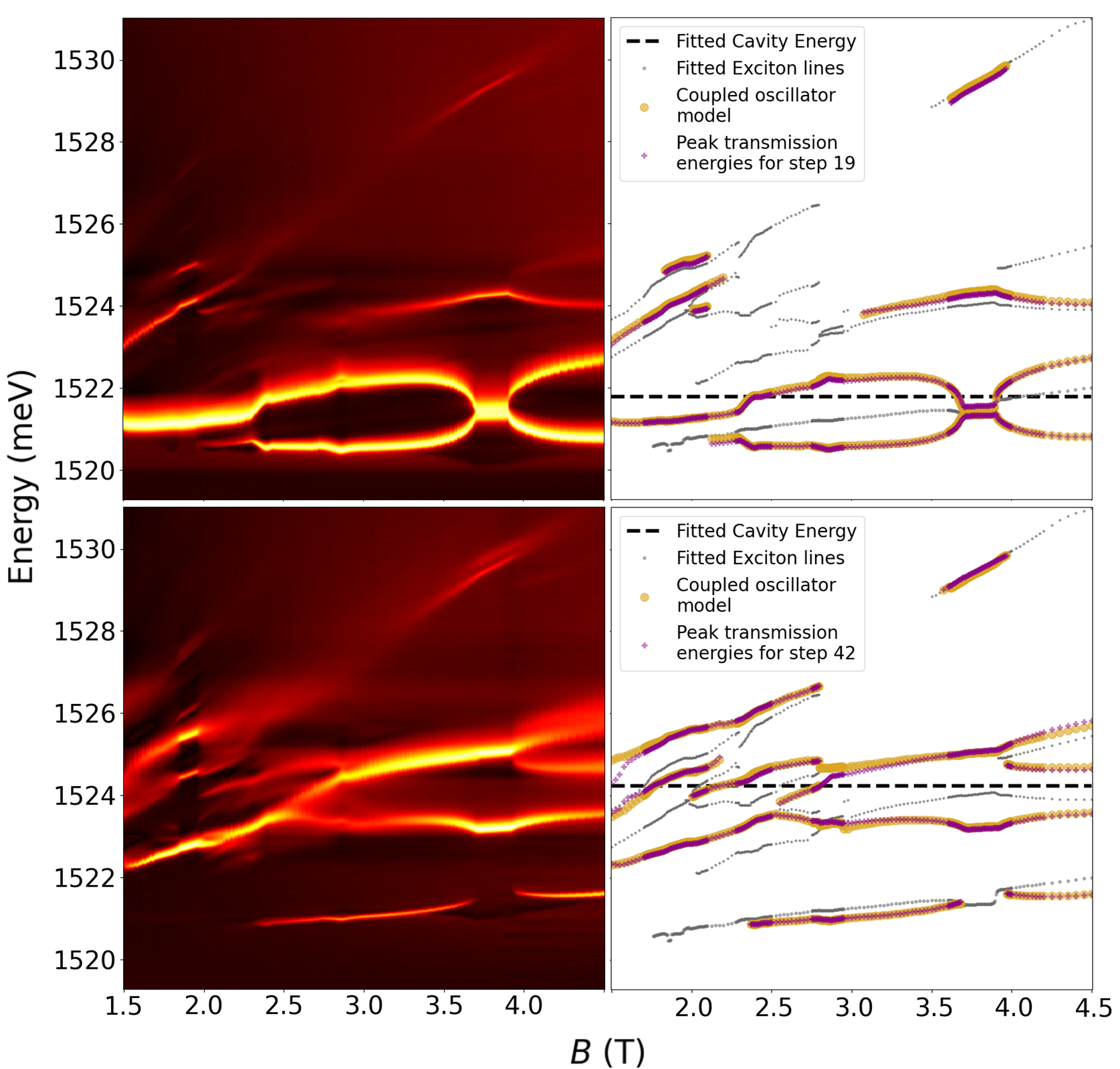}
    \caption{Left: Differential reflection spectra for positional step 19 (top) and 42 (bottom) of $\sigma_-$ measurement set in sample A. Right: Extracted transmission peaks (purple points) and results from fitting procedure, where the black dashed lines represent the cavity energy, the grey points are the excitation energies, and the golden points are the theoretical polariton energies we fit to our data.}\label{fig:raw_data_example}  
\end{figure}

\begin{figure}[htb]  
    \includegraphics[width=3.4in]{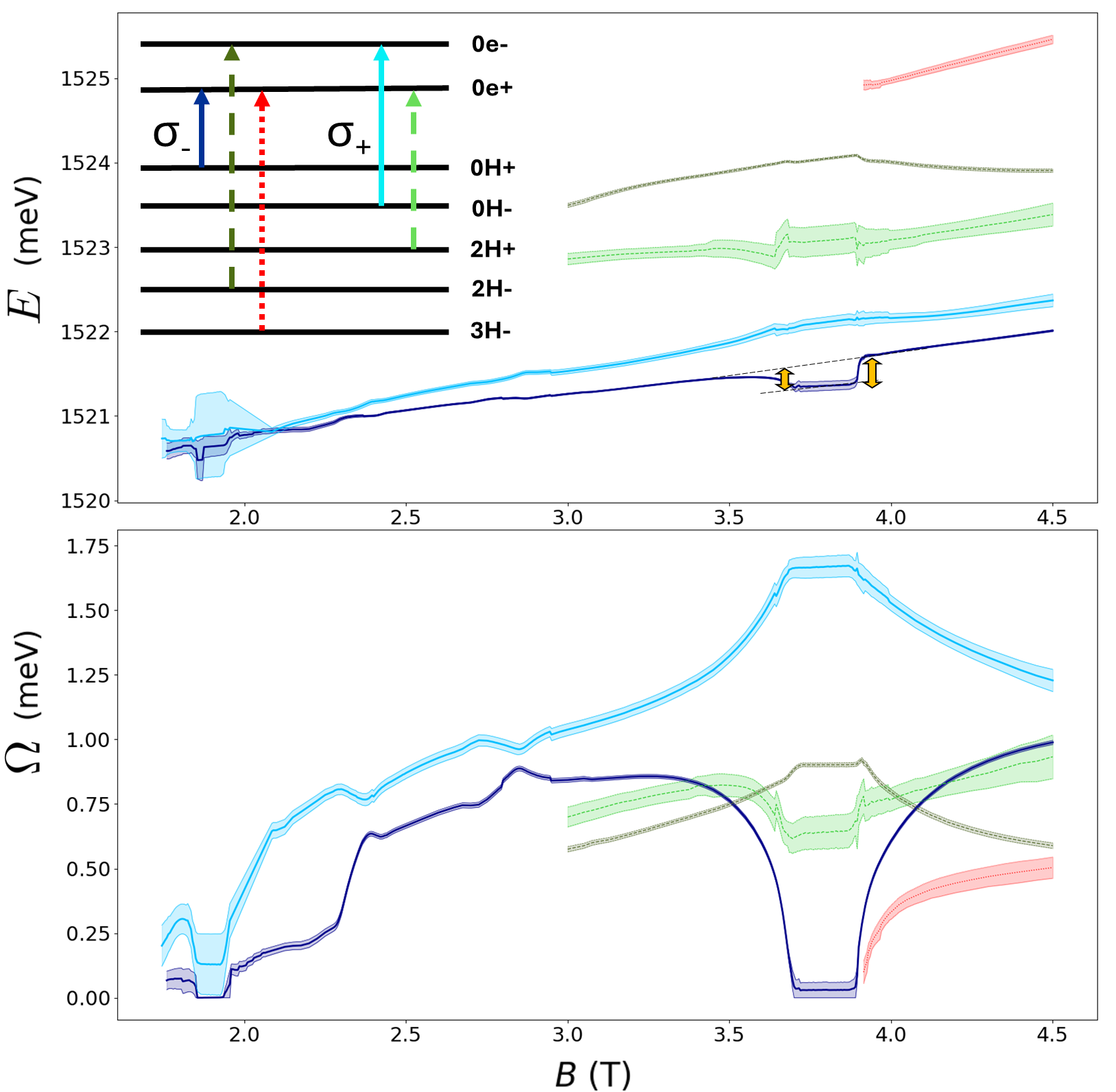}
    \caption{Optical excitation energies (top) and Rabi coupling (bottom) of sample A as a function of $B$ field. Dark blue, dark green and red transistions are extracted from $\sigma_-$ data set, while light blue and light green are from $\sigma_+$ set. The inset in the top left corner indicates the transitions the different lines correspond to. The yellow arrows signal a jump in the lowest energy excitation happening close to $\nu=1$.}\label{fig:optical_transitions}
\end{figure}

Following the discussion from Section \ref{sec:experiment}, the transition energies to the different LLs and their Rabi coupling as a function of $B$ field could be measured. $B$ field sweeps such as the one shown in Figure \ref{fig:large_B_sweep} were performed, but for multiple positions $x$ across the sample. This yielded $n$ sets of polariton energies $\mathcal{P}_0(B,x),\dots,\mathcal{P}_n(B,x)$. To this we fitted a $n\times n$ coupled oscillator Hamiltonian \ref{eq:coupled_oscillator_ham}, where the $E(B)$ and $\Omega(B)$ parameters (corresponding to the LLs and their coupling) were arbitrary functions of $B$ (i.e. 1 fit parameter per $B$ value measured) and $E_C(x)$ was a second order polynomial, in analogy with the measurement presented in Section \ref{sec:cavity_tuning}.

Figure \ref{fig:raw_data_example} illustrates the fitting procedure for sample A under $\sigma_-$ circular polarization, focusing on the range $2>\nu>1$. The top data set has a cavity energy tuned close to the $E_0$ transition at $\nu=1$, while the bottom data set is tuned near $E_1$ at $\nu=1$ (the nature of $E_0$ and $E_1$ will be discussed later). 

The low-energy dataset (top) reproduces the discussion of section \ref{subsec:LL pol sig}: the lowest polariton line appears at $B=1.9$ T ($\nu=2$), and collapses to 0 splitting at $\nu=1$. Our fitting procedure accurately captures these features. The high-energy dataset (bottom) exhibits stronger resonance with higher-energy transitions, resulting in more complex spectra with overlapping transitions and broadened linewidths. While our fitting procedure remains effective at these higher energies, it's important to note that the extracted data points may not fully represent the complete spectral features due to the rejection of faint transmission peaks.

As the polariton order increases, the accuracy of the fit values decreases, both in terms of error value and intrinsic model accuracy. Consequently, our subsequent analysis primarily focuses on the lowest energy transition.

Figure \ref{fig:optical_transitions} presents the extracted fit values for sample A for both optical circular polarizations, showing optical excitation energies and Rabi coupling as a function of B field. For clarity, we've omitted higher energy transitions deemed less accurate. The two lowest transitions correspond to 0\textbf{H}$^{+}\rightarrow$ 0\textbf{e}$^{+}$ ($\sigma_-$ polarization) and 0\textbf{H}$^{-}\rightarrow$ 0\textbf{e}$^{-}$ ($\sigma_+$ polarization), where 0 stands for $L=0$, \textbf{e} and \textbf{H} are for conduction (electron) or valence (hole) bands, and + or - corresponds to the spin state \cite{Rav2018,Aifer1996}. 

As the holes in the valence band have a much larger effective mass than electrons in the conduction band, the variation of the hole cyclotron energy as a function of $B$ is negligible compared to the electron variation. Neglecting all other band bending effects, we can estimate the electron effective mass from the average slope of these two lowest energy transitions: $m^{*}=(0.105\pm0.002)m_e$, with $m_e$ the free electron mass. This value is consistent with previous results, supporting the accuracy of our fit. \cite{Haldar2017}.

The higher energy transitions have similar $B$ field dependencies. As such, they should also correspond to transitions to 0\textbf{e}$\pm$ states. Indeed, the 2\textbf{H}$^{-}\rightarrow$ 0\textbf{e}$^{-}$ and 3\textbf{H}$^{-}\rightarrow$ 0\textbf{e}$^{+}$ in $\sigma_-$ polarization have been previously measured and theorised \cite{Aifer1996}. Their Rabi coupling is also consistent with this interpretation, where transitions to the 0\textbf{e}$^{+}$ state see their coupling vanish as $\nu=1$ is approached, while transitions to the 0\textbf{e}$^{-}$ state have maximal coupling at $\nu=1$.

The light green transition from the $\sigma_+$ resonance was however not explained in the literature. Reference \cite{Aifer1996} has data and calculations for the other four transitions. Their data also appears to have a faint second peak in $\sigma_+$ polarization which they do not discuss. We tentatively label it here as 2\textbf{H}$^{+}\rightarrow$ 0\textbf{e}$^{+}$ out of analogy with the dark green transition, but accurate calculations are still needed to confirm this interpretation. Again, although not shown here, samples B and C exhibited the same number of transitions with the same Rabi coupling variation for their two circular polarizations.

Finally, we would like to note that the energy of the 0\textbf{H}$^{+}\rightarrow$ 0\textbf{e}$^{+}$ transition in sample A exhibits a small dip around $B=3.8$ T, signalled by yellow arrows in Figure \ref{fig:optical_transitions}. Although the energy measurement at this point is inaccurate due to the vanishing optical coupling, the low error bars from our model fit suggest this effect is real. We measure the step size to be  $\Delta_{\nu>1}=0.1$ meV and $\Delta_{\nu<1}=0.35$ meV. This feature can be related to the compressibility of the 2DEG. The energy $E$ measured on either side of $\nu=1$ is the chemical potential $\mu$ of the $e$-$h$ pair, with $\mu = E_G + \Delta$, where $E_G$ is the ground state energy of the 2DEG and $\Delta$ is the additional energy required to compress the system. As the system becomes incompressible at $\nu=1$, $\Delta\rightarrow\infty$ and the optical coupling vanishes. The very small finite coupling we measure in our data can be explained by sparse vacancies in the 2DEG owing to the finite temperature of the system. In this state, exciting electrons would only require an energy $E_G$, as we would not be compressing the system but only filling in temperature-dependent vacancies.

\subsection{2DEG spin polarization: $\nu = 1$ Quantum Hall Ferromagnet and $\nu=3/2$ CF effective mass}
\label{subsec:spin_polarization}

The spin polarization of the 2DEG can be deduced from the ratios of coupling strengths of the two opposite spin lowest energy transitions (dark and light blue lines in Figure \ref{fig:optical_transitions}). The square of the coupling strength is proportional to the density of empty states in the excited level: $\Omega^2 = C\rho$ with $C$ the proportionality constant and $\rho$ the density of empty states \cite{Aifer1996,Manfra1996}. If we assume $C$ to be the same for the $\sigma_-$: 0\textbf{H}$^{+}$ $\rightarrow$ 0\textbf{e}$^{+}$ and $\sigma_+$: 0\textbf{H}$^{-}\rightarrow$ 0\textbf{e}$^{-}$ transitions then we get
\begin{equation}
\label{eq:optical_pol}
P_{\Omega}=\frac{\Omega_{\sigma_+}^2 - \Omega_{\sigma_-}^2}{\Omega_{\sigma_+}^2 + \Omega_{\sigma_-}^2} = \frac{\rho_{-} - \rho_{+}}{\rho_{-} + \rho_{+}} = \frac{n_{+} - n_{-}}{2n_{LL} - n_{+} - n_{-}}
\end{equation}
with the three terms $n_{LL} = n_{+(-)} + \rho_{+(-)}$ designating respectively the density of degeneracy, occupied and empty states of the 0\textbf{e}+($-$) energy level. For $\nu<2$, $n_{LL} = n_e/\nu$, and $n_{+} + n_{-} = n_e$ so Eq. \ref{eq:optical_pol} becomes the normalized spin polarization of the system:
\begin{equation}
P_{\Omega} = \frac{\nu}{2-\nu}P_{S}
\end{equation}
with $P_{S} = (n_+ - n_-)/n_e$ the spin polarization.

\begin{figure}[htb]  
    \includegraphics[width=3.4in]{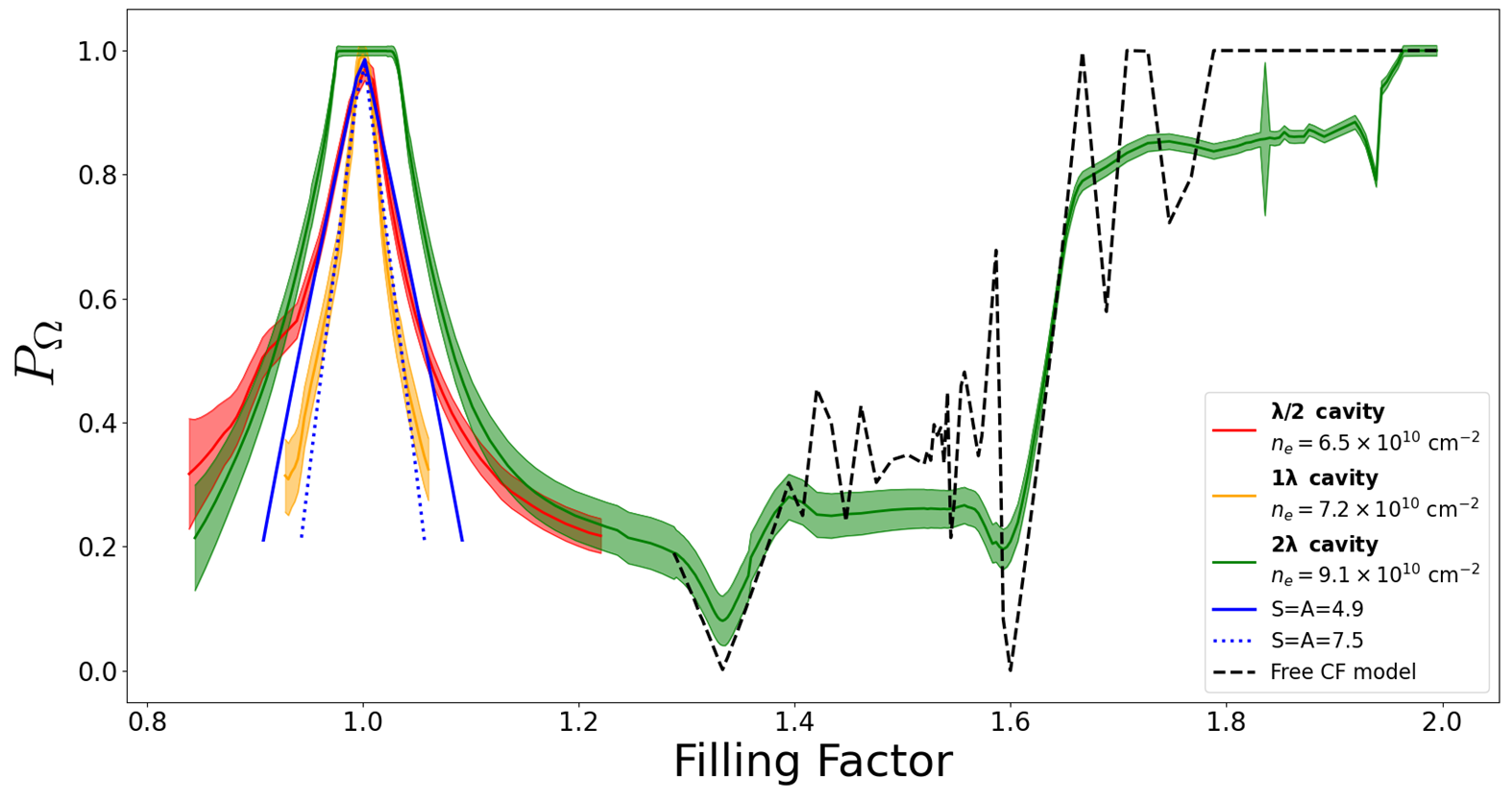}
    \caption{Normalized spin polarization of samples A (green), B (yellow) and C (red) as a function of filling factor. Skyrmion models with a single skyrmion size fit parameter were fitted to the B data (blue dotted line) and C data (blue full line). The black dotted line is obtained from a CF model with no fit parameters.}\label{fig:spin_polarization}

\end{figure}

Figure \ref{fig:spin_polarization} shows the normalized spin polarization of our three samples as a function of their filling factor. At $\nu=1$, all three samples exhibit full spin polarization, corresponding to a ferromagnetic state, where all spins are aligned. This $\nu=1$ Quantum Hall Ferromagnet is a widely reported result, typically accompanied by rapid depolarization on either side of $\nu=1$, attributed to Skyrmion and Anti-Skyrmion formation. Samples B and C display this characteristic behaviour. The spin polarization in this regime is given by \cite{Aifer1996}
\begin{align}
P_{S} = \frac{1}{\nu} - (2A-1)\frac{1-\nu}{\nu},\;\;\unit{for}\;&\nu<1\\
P_{S} = S\left(\frac{2-\nu}{\nu}-1\right) + 1,\;\;\unit{for}\;&\nu>1
\end{align}
With $S$ ($A$) the number of electron spins flipped per Skyrmion (Anti-Skyrmion). We assume $S=A$, based on symmetry arguments. The fitted $S=A$ parameter for samples B and C is 7.5 and 4.9 respectively, which is within the range of previously observed values \cite{Aifer1996,Manfra1996,Lupatini2020,Plochoka2009}. The Skyrmion size is correlated to the ratio $E_X/E_Z$ (exchange and Zeeman energies). At $\nu=1$, $E_{Z,B}>E_{Z,C}$, owing to the larger $n_e$ of sample B. We can conclude the exchange energy in sample B is larger, which can be interpreted as a longer coherence length of the 2DEG \cite{Sinova2000}. We attribute the shorter coherence length of sample C to increased remote impurity scattering due to the AlAs layer in close proximity to the QW.

Sample A, however, exhibits an unusually robust ferromagnet, maintaining full polarization in the range $0.98<\nu<1.03$. This persistence of ferromagnetism at the $\nu=1$ state has not been reported before. Intriguingly, the coupling $\Omega_{\sigma_-}$ vanishes at values lower than $\nu=1$, which should not be possible since the 0\textbf{e}$^{+}$ ground state cannot be completely full at this point. 

While the underlying mechanism remains unclear, we hypothesize a connection to the incompressibility of this state, remarked upon in Section \ref{subsec:optical transitions}. A previous study has measured an enhanced nonlinear optical response of the $\nu =$ 1, 2/3 and 2/5 states \cite{Knueppel2019}. It is possible that a large amount of polaritons in the system have made it reach its incompressible limit, thereby extending the size of the vanished optical coupling plateau. This discussion will be extended in Section \ref{subsec:power_temp_dep}.

We would like to conclude this section with the spin polarization for $\nu>1.2$. In-depth analysis was only carried out for sample A, but measurements of samples B and C observed qualitative agreement. These results were compared to a model attributing an effective mass of $m^{*}_{3/2} = 0.45 m_e$ to the $\nu=3/2$ CF, allowing to make calculations based on the ratio of Zeeman to cyclotron energy affecting the CF around this filling factor \cite{Park1998}, assuming no disorder and no CF interactions. As shown in Figure \ref{fig:spin_polarization}, both data and model agree on a depolarization of the 2DEG at $\nu =$ 4/3 and 8/5, followed by strong repolarization at $\nu=5/3$. Maximal normalized polarization as $\nu=2$ is approached then corresponds to a Wigner crystal phase \cite{Yoo2020}.

\subsection{Optical power and temperature dependence of the $\nu=1$ ferromagnetic plateau}
\label{subsec:power_temp_dep}

The vanishing Rabi coupling of sample A was investigated further with temperature and optical power. As increasing the power output of the LED inherently heats up the system, we first studied the isolated effect of temperature. 

\begin{figure}[htb]  
    \includegraphics[width=3.4in]{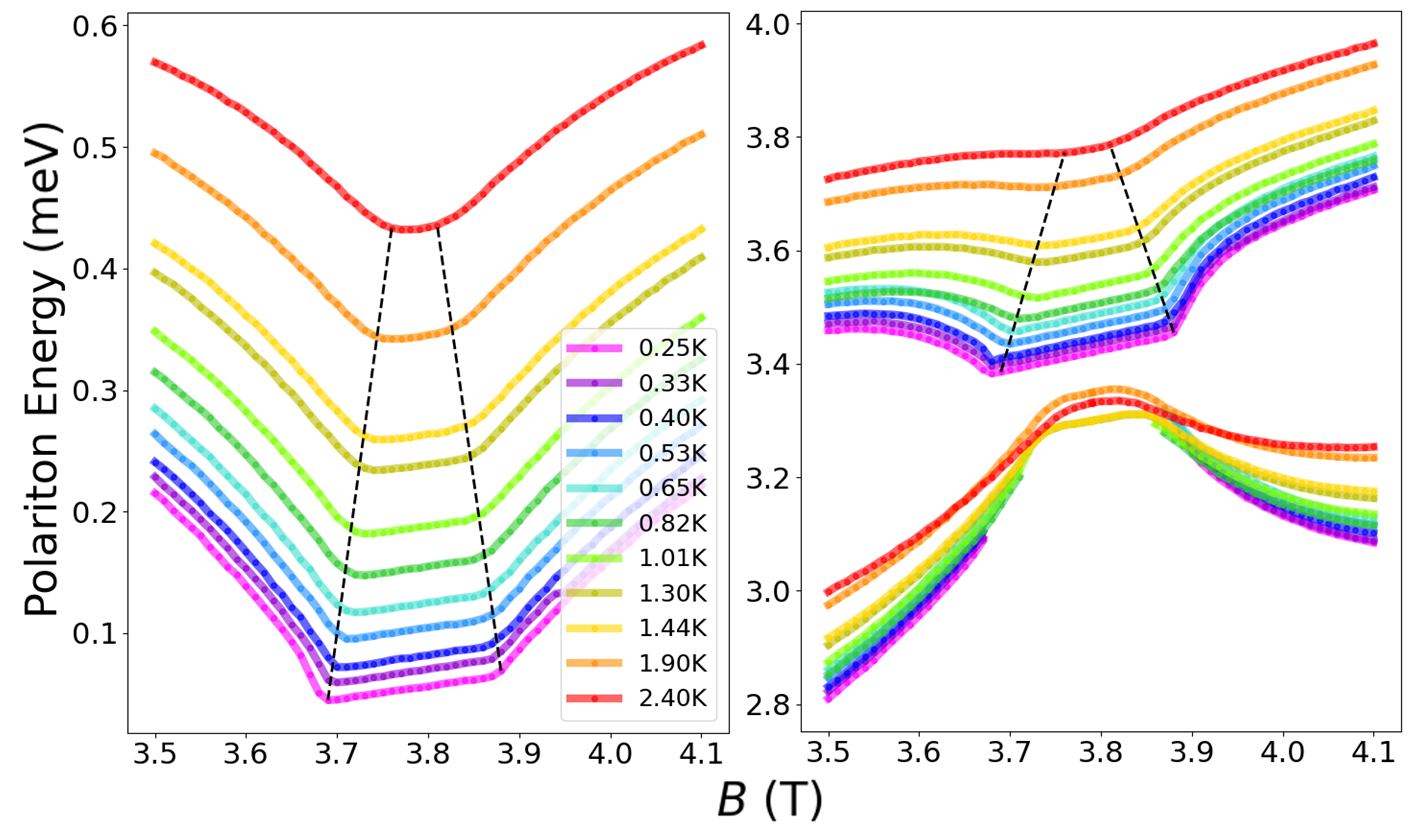}
    \caption{Polariton energies in $\sigma_+$ polarization (Left: $\mathcal{P}_0$ , Right: $\mathcal{P}_1$ and $\mathcal{P}_2$) close to $\nu=1$ with fixed cavity energy $E_C = 1521$ meV. The energies are shifted such that the minimum energy of $\mathcal{P}_0$ is proportional to the measurement temperature, shown in the inset. In absolute terms, all the minimum energies coincided at 1519.95 meV. Black dashed lines guide the eye, showing the evolution of the $\nu=1$ plateau.}\label{fig:temp_dep}

\end{figure}

Figure \ref{fig:temp_dep} shows the evolution of the $\nu=1$ state for temperatures in the range $T = 0.25$ K to $T=2.4$ K, at fixed cavity energy $E_C = 1521$ meV. In $\sigma_+$ polarization here, we know, according to Figure \ref{fig:optical_transitions}, that the three polariton branches $\mathcal{P}_0$, $\mathcal{P}_1$ and $\mathcal{P}_2$ emerge from the cavity coupling to the transitions 0: 0\textbf{H}$^{-}\rightarrow$0\textbf{e}$^{-}$ and 1: 2\textbf{H}$^{+}\rightarrow$ 0\textbf{e}$^{+}$. $\mathcal{P}_0$ and $\mathcal{P}_2$ are split further apart both when $\Omega_0$ and $\Omega_1$ are increased, since both these transition energies sit in between the branches. $\mathcal{P}_1$ however sits in between transitions 0 and 1, meaning it is pushed up when $\Omega_0$ is increased but pushed down when $\Omega_1$ is increased. Thus, the energy of $\mathcal{P}_1$ correlates to $\Omega_0^2-\Omega_1^2$, while the separation of $\mathcal{P}_0$ and $\mathcal{P}_2$ correlates to $\Omega_0^2+\Omega_1^2$. 

It should be noted that although the coupling $\Omega_1$ does not vanish at $\nu=1$ in Figure \ref{fig:optical_transitions}, we consider this to be due to the lack of measurement of higher energy polaritons in this polarization, which may have led the model to compensate their unaccounted coupling with increased coupling of this transition. Key observations from Figure \ref{fig:temp_dep} include:
\begin{enumerate}
    \item The disappearance of $\mathcal{P}_1$ at $\nu=1$ at low temperatures suggests $\Omega_1$ is very small or 0. 
    \item The overall coupling $\Omega_0^2 + \Omega_1^2$ appears temperature-independant, as $\mathcal{P}_0$ and $\mathcal{P}_2$ maintain similar energies across temperatures. 
    \item The energy variation of $\mathcal{P}_1$ decreases as the temperature is increased, indicating slower depolarization of the 2DEG away from $\nu=1$. This is consistent with the Skyrmion size $S$ decreasing with increased temperature \cite{Plochoka2009}. 
    \item The re-emergences of $\mathcal{P}_1$ at $\nu=1$ for $T>1.3$ K indicates the 2DEG is no longer fully polarized at these temperatures. 
    \item The full polarization plateau narrows with increasing temperature: 
    \begin{enumerate}
        \item At $T=0.25$ K: $3.69\,\unit{T}<B<3.88\,\unit{T}$ ($1.03>\nu>0.98$)
        \item At $T=2.4$ K:  $3.76\,\unit{T}<B<3.81\,\unit{T}$ ($1.01>\nu\ge1$).
    \end{enumerate}
\end{enumerate}
The linear appearance of the black dashed lines suggests a possible linear relationship between plateau size and temperature.

\begin{figure}[htb]  
    \includegraphics[width=3.4in]{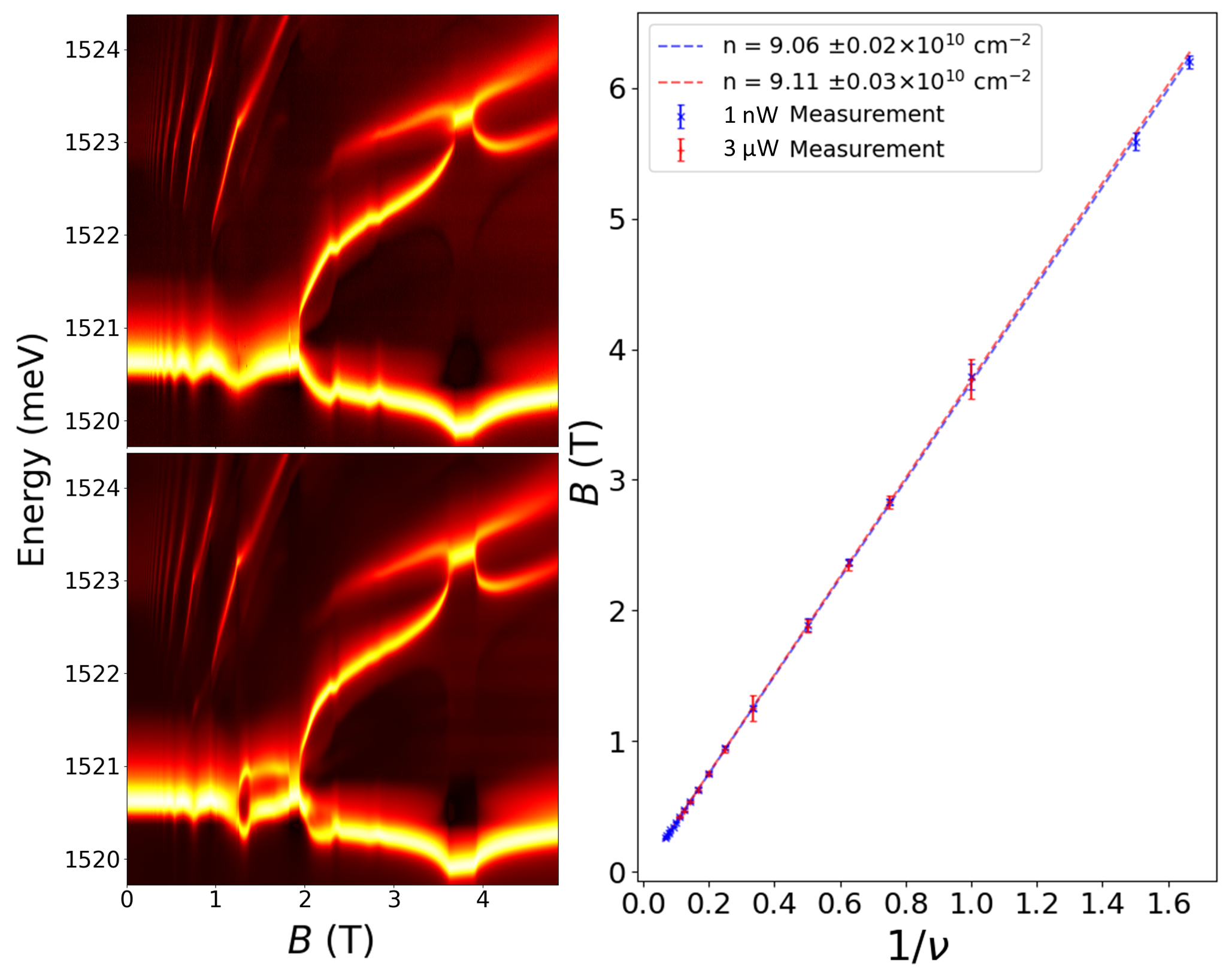}
    \caption{Left: Differential reflection spectra of sample A in $\sigma_+$ polarization with top: 1 nW LED power, or bottom: 3 $\mu$W LED power, with a cavity energy fixed close to $E_C = 1521$ meV. Right: $B$ values of maxima and minima of lowest polariton line corresponding to different filling factors $\nu$, used to measure the 2DEG density with $B = n_e h/e\nu$.}\label{fig:power_density}

\end{figure}

To verify the effect of optical power on the coupling strength at $\nu=1$, we first ensured that the 2DEG density remained constant with changing power, unlike in many semiconductor devices. Figure \ref{fig:power_density} shows 2 sets of transmission spectra, with the bottom set being made with 3000 times greater optical power. The main differences observed between the two sets are a widening of the $\nu=1$ ($B=3.8$ T) ferromagnetic plateau at higher power, as well as a strange re-emergence of the middle polariton line for $1.27\,\unit{T}<B<1.9\,\unit{T}$ ($3>\nu>2$). Other than that, the lower polariton line has a broader linewidth and the higher order filling factors are not as well defined at higher power, which was to be expected. It can be however stated that there is no measurable change in 2DEG density at these different powers.

In contrast, the 2DEG density $n_e$ of samples B and C decreased by $-0.2\times10^{10}$cm$^{-2}$ every time the optical power density $W$ was multiplied by 10. This effect is attributed to the formation of DX centers due to the AlAs barrier adjacent to the doping QWs in these devices \cite{Miwa1999,Ababou1990}. This light sensitivity may be the reason a large plateau of vanished coupling was only observed for sample A.

\begin{figure}[htb]  
    \includegraphics[width=3.4in]{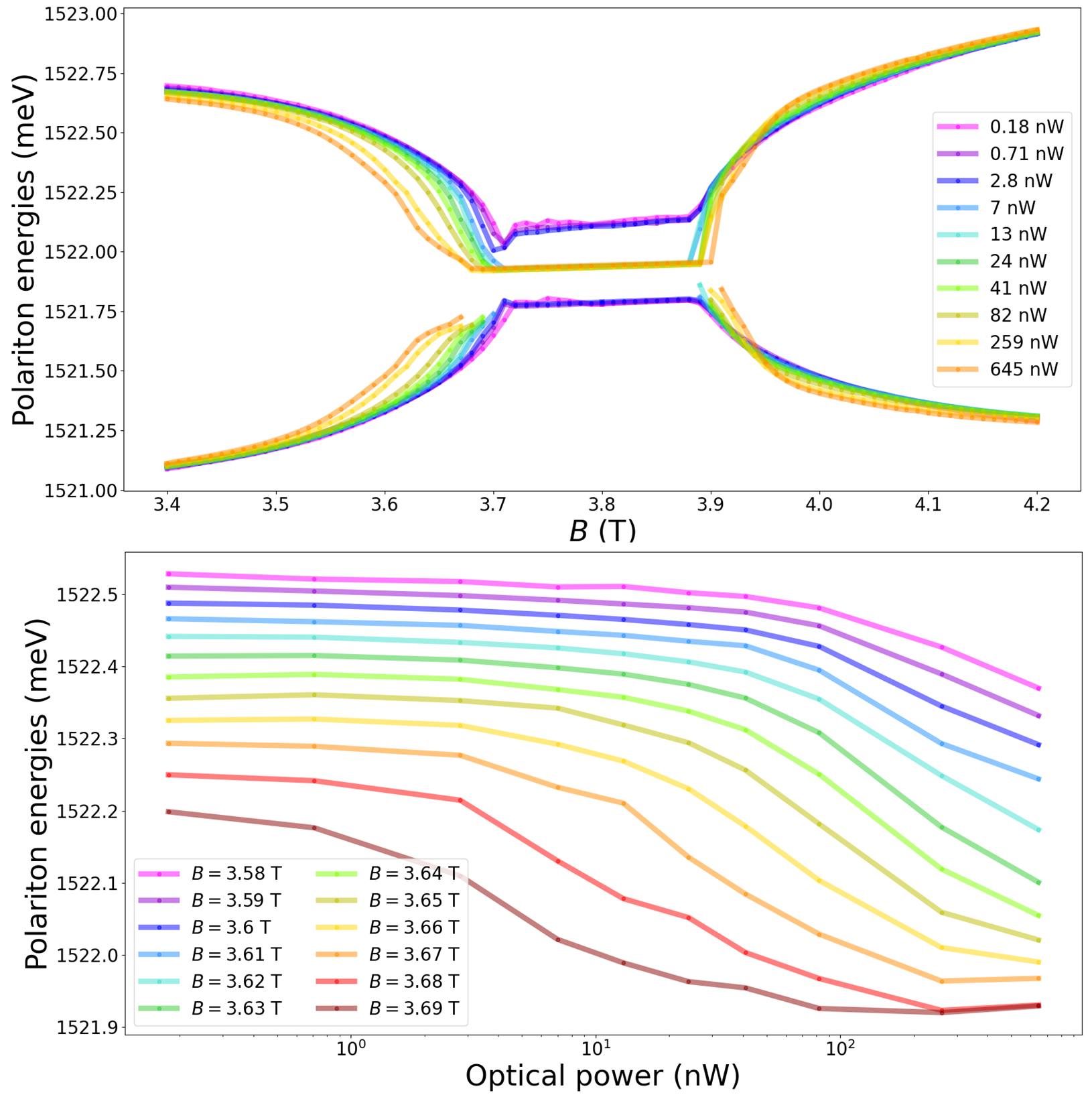}
    \caption{Top: Polaritons $\mathcal{P}_0$ and $\mathcal{P}_1$ in $\sigma_-$ polarization, at fixed cavity energy $E_C=1522$ meV around $\nu=1$, for different LED powers, shown in the inset. Bottom: Energy of polariton 1 as a function of optical power at different $B$ fields, shown in the inset.}\label{fig:power_plateau}
\end{figure}

Figure \ref{fig:power_plateau} shows the evolution of the $\nu=1$ plateau with optical power in sample A. A direct measurement of the optical power density $W$ at the focal point could not be performed, but its value should be proportional to the optical power shown here. 

Increasing the power by 3 orders of magnitude resulted in a widening of the plateau size, in stark contrast with Figure \ref{fig:temp_dep} showing narrowing of the plateau with increased temperature. At $B=3.4$ T and $B=4.2$ T, the polariton energies are unaffected by the optical power, indicating this effect is strictly limited to the $\nu=1$ plateau. One other feature of this measurement is that two distinct polariton lines reappear for powers less than 1 nW over the range of the $\nu=1$ plateau (dark blue and violet lines), instead of a single line of vanished coupling. This marks a transition from the weak to the strong coupling regime, where the low power polariton lifetime is large enough to resolve the two polaritons.

Increasing the power to greater values (not shown here) then produced the opposite effect, with the plateau size decreasing. However, further modification of the polariton energy over the whole range $3.4\unit{T} < B<4.2\unit{T}$ in this regime suggest a thermal limit was crossed, where the 2DEG is locally warmed up by the high power LED.

\section{Conclusion and outlook}

In conclusion, we have demonstrated the strength of monolithic GaAs-based microcavities for the exploration of quantum Hall physics. Although the results here focus on sample A, qualitative agreement from samples B and C emphasize the robustness of our design. The complete light insensitivity of the 2DEG density exclusive to sample A and the reduced Skyrmion size of sample C propone the 2$\lambda$ cavity design as the best approach of the three.

The relatively large Skyrmion sizes of samples B and C, as well as the good agreement with the free CF model of sample A, from the spin polarization measurements, bring further confirmation to our understanding of IQH and FQH states in the range $\nu<2$. They also serve as a testament to the low amount of perturbation produced by the measurement. As such, we would like to reaffirm that studying more exotic FQH states through this approach is within reach, with two obvious improvements being lower measurement temperature and larger 2DEG densities.

Finally, the large range of vanished coupling at $\nu=1$ remains without formal explanation, although the opposition of observed effects for increased optical power or sample temperature suggests we have reached a non-linear optics regime.

\begin{acknowledgements}
We thank Martin Kroner and Patrick Kn\"uppel for insightful discussions, both in terms of the physics discussed here as well as the experimental realization of these measurements. Also extended thanks to the rest of the Wegscheider group for their technical assistance, in particular Christian Reichl and Amina Ribero for their help with our $^{3}$He cryostat. We acknowledge financial support from the Swiss National Science Foundation (SNSF) and the NCCR QSIT (National Center of Competence in Research - Quantum Science and Technology).
\end{acknowledgements}

\bibliography{bibliography}

\end{document}